\documentclass[a4paper,12pt,nonacm]{acmart}
\usepackage{graphicx}
\usepackage{tikz}
\usepackage{geometry}
\usepackage{xcolor}
\usepackage[T1]{fontenc}
\usepackage{tgbonum}
\usepackage{lmodern}
\usepackage{tgtermes}
\usepackage{tgpagella}
\usepackage{tgschola}
\usepackage{mathptmx}
\usepackage{utopia}
\usepackage{palatino}
\usepackage{bookman}
\usepackage{charter}
\usepackage{multicol}
\usepackage{underscore}
\usepackage{float}
\usepackage{placeins}
\usepackage{wrapfig} 
\usepackage{enumitem}
\usepackage{array}
\usepackage{hyperref}
\hypersetup{colorlinks=true, linkcolor=blue, urlcolor=blue}
\definecolor{urlblue}{RGB}{0,0,238} 

\usepackage[resetlabels,labeled]{multibib}
\newcites{Memo}{Further Readings}

\setitemize{noitemsep,topsep=0pt}


\definecolor{darkgreen}{RGB}{0,100,0}
\newcommand{\inlinecode}[1]{\textcolor{darkgreen}{\textbf{\texttt{#1}}}}

\definecolor{offwhite}{RGB}{220,220,220}

\newcommand{\papertitle}{Unearthing a Billion Telegram Posts about the 2024 U.S. Presidential Election: Development of a Public Dataset}

\newcommand{\paperauthors}{Leonardo Blas, Luca Luceri, Emilio Ferrara}
\newcommand{\paperaffiliation}{University of Southern California} 

\begin{document}
	
	\begin{titlepage}
		\begin{tikzpicture}[remember picture, overlay]
			\node[anchor=north west, inner sep=0] at (current page.north west) {
				\includegraphics[width=\paperwidth, height=\paperheight, trim=275 375 275 375]{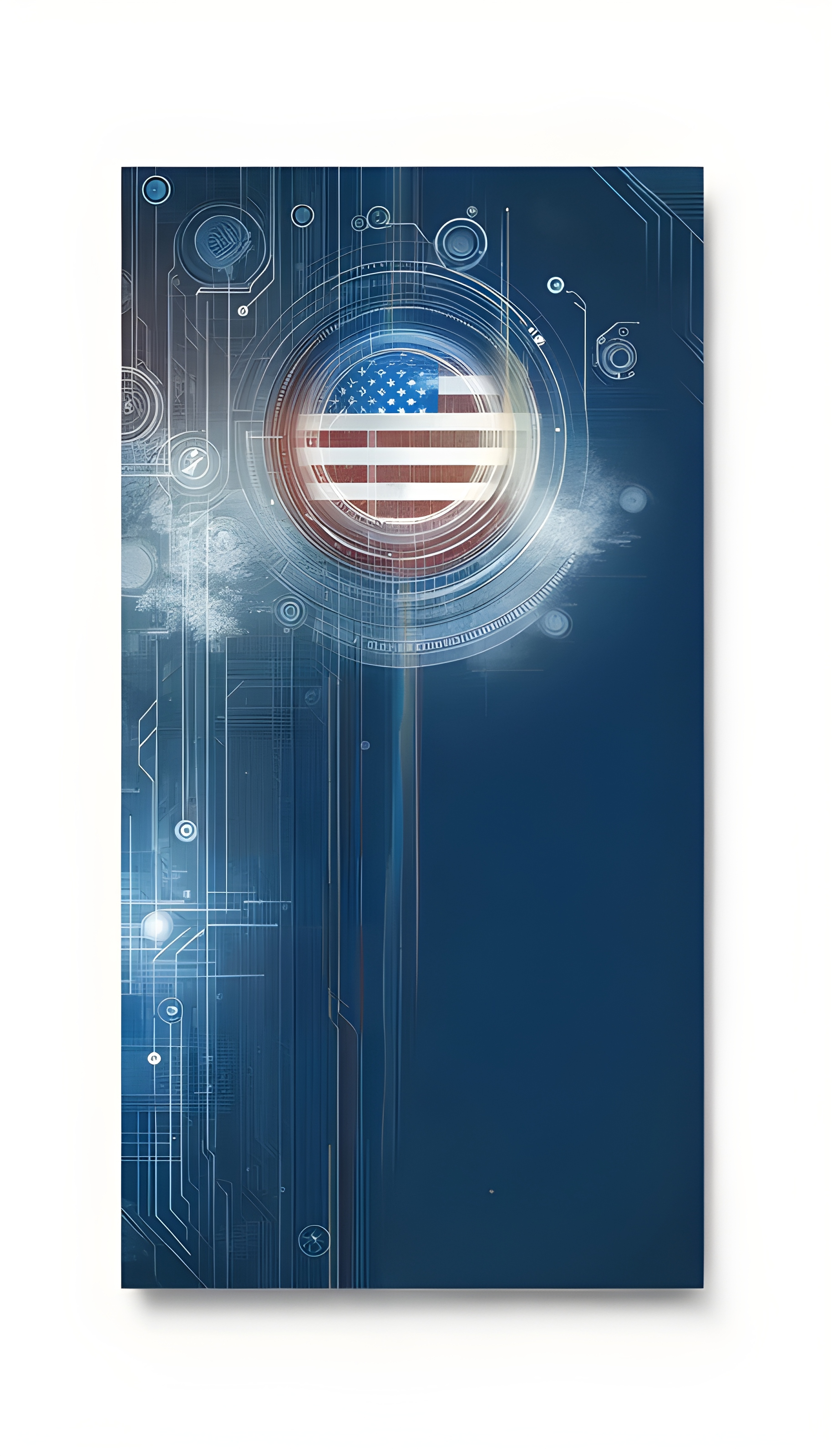}
			};
			\node[anchor=center, yshift=-9cm] at (current page.center) {
				\begin{minipage}{\textwidth}
					\raggedleft
					\color{offwhite}
					{\Huge \bfseries \fontfamily{qtm}\selectfont The 2024 Election Integrity Initiative }
					
					\vspace{1.5cm}
					
					{\LARGE \fontfamily{qtm}\selectfont \papertitle}
					
					\vspace{1.5cm}

					{\Large \fontfamily{qtm}\selectfont \paperauthors}
					
					\vspace{1cm}
					
					{\Large \fontfamily{qtm}\selectfont \paperaffiliation}
					
					\vfill
					
					{\Large \fontfamily{qtm}\selectfont HUMANS Lab -- Working Paper No. 2024.5}
				\end{minipage}
			};
		\end{tikzpicture}
	\end{titlepage}
	
	\noindent{\LARGE \fontfamily{qtm}\selectfont \papertitle}
	
	\vspace{0.5cm}
	
	\noindent{\large \fontfamily{qtm}\selectfont \paperauthors}
	
	\noindent{\large \fontfamily{qtm}\selectfont \textit{\paperaffiliation}}
	
	\section*{Abstract}

With its lenient moderation policies and long-standing associations with potentially unlawful activities, Telegram has become an incubator for problematic content, frequently featuring conspiratorial, hyper-partisan, and fringe narratives. In the political sphere, these concerns are amplified by reports of Telegram channels being used to organize violent acts, such as those that occurred during the Capitol Hill attack on January 6, 2021. As the 2024 U.S. election approaches, Telegram remains a focal arena for societal and political discourse, warranting close attention from the research community, regulators, and the media. Based on these premises, we introduce and release a Telegram dataset focused on the 2024 U.S. Presidential Election, featuring over 30,000 chats and half a billion messages, including chat details, profile pictures, messages, and user information. We constructed a network of chats and analyzed the 500 most central ones, examining their shared messages. This resource represents the largest public Telegram dataset to date, offering an unprecedented opportunity to study political discussion on Telegram in the lead-up to the 2024 U.S. election. We will continue to collect data until the end of 2024, and routinely update the dataset released at:  \url{https://github.com/leonardo-blas/usc-tg-24-us-election}

	\section*{Introduction}

In recent years, Telegram has emerged as one of the most widely used messaging platforms globally, boasting a diverse array of communities that span personal, social, and political interests. Despite its popularity, Telegram has gained a reputation for lenient moderation policies, which, alongside long-standing associations with potentially unlawful activities, have positioned the platform as an incubator for problematic content~\cite{simon2023linked}. Frequently, Telegram channels and groups host discussions that feature conspiratorial, hyper-partisan, and fringe narratives~\cite{hoseini2024characterizing}. This aspect of Telegram's ecosystem has raised concerns within the broader social and political landscape, as researchers, policymakers, and media outlets increasingly question the platform’s role in amplifying controversial viewpoints and influencing public discourse~\cite{junior2021towards}.

The political implications of Telegram’s role are particularly notable. Past reports suggest that Telegram has been used as an organizing tool for coordinated and sometimes violent actions~\cite{walther2021us, bovet2022organization}, as exemplified by its role in the events leading up to the Capitol Hill attack on January 6, 2021~\cite{scheffler2021telegram}. This incident underscored the platform’s potential to facilitate real-world mobilization among like-minded users, a risk that becomes especially relevant as the 2024 U.S. election approaches. Given the heightened stakes of this election, Telegram has once again become a focal arena for societal and political discussions, creating a need for systematic monitoring and analysis of its content.
In light of these concerns, this paper introduces and publicly releases a large-scale dataset of Telegram discussions focused on the 2024 U.S. Presidential Election. Our dataset comprises over 30,000 chats and over half a billion messages, to date, providing comprehensive data, including chat details, profile pictures, messages, and user information. This dataset also enables the construction of a network of chats, facilitating the analysis of interconnected communities and the content shared within them. Here, we specifically focus on the 500 most central chats in this network, examining the messages exchanged in these influential hubs to understand how political discourse is shaped and circulated within Telegram.

This resource represents the largest publicly available Telegram dataset to date, to the best of our knowledge, offering an unprecedented opportunity to study the dynamics of political discussions on Telegram in the run-up to the 2024 U.S. election. By making this dataset available, we aim to provide researchers, policymakers, and media with valuable insights into Telegram's role in shaping public discourse and influencing political attitudes, ultimately contributing to a better understanding of the platform’s impact on democratic discussion.
 

	\begin{wrapfigure}{R}{0.35\textwidth}
		\includegraphics[width=0.35\textwidth]{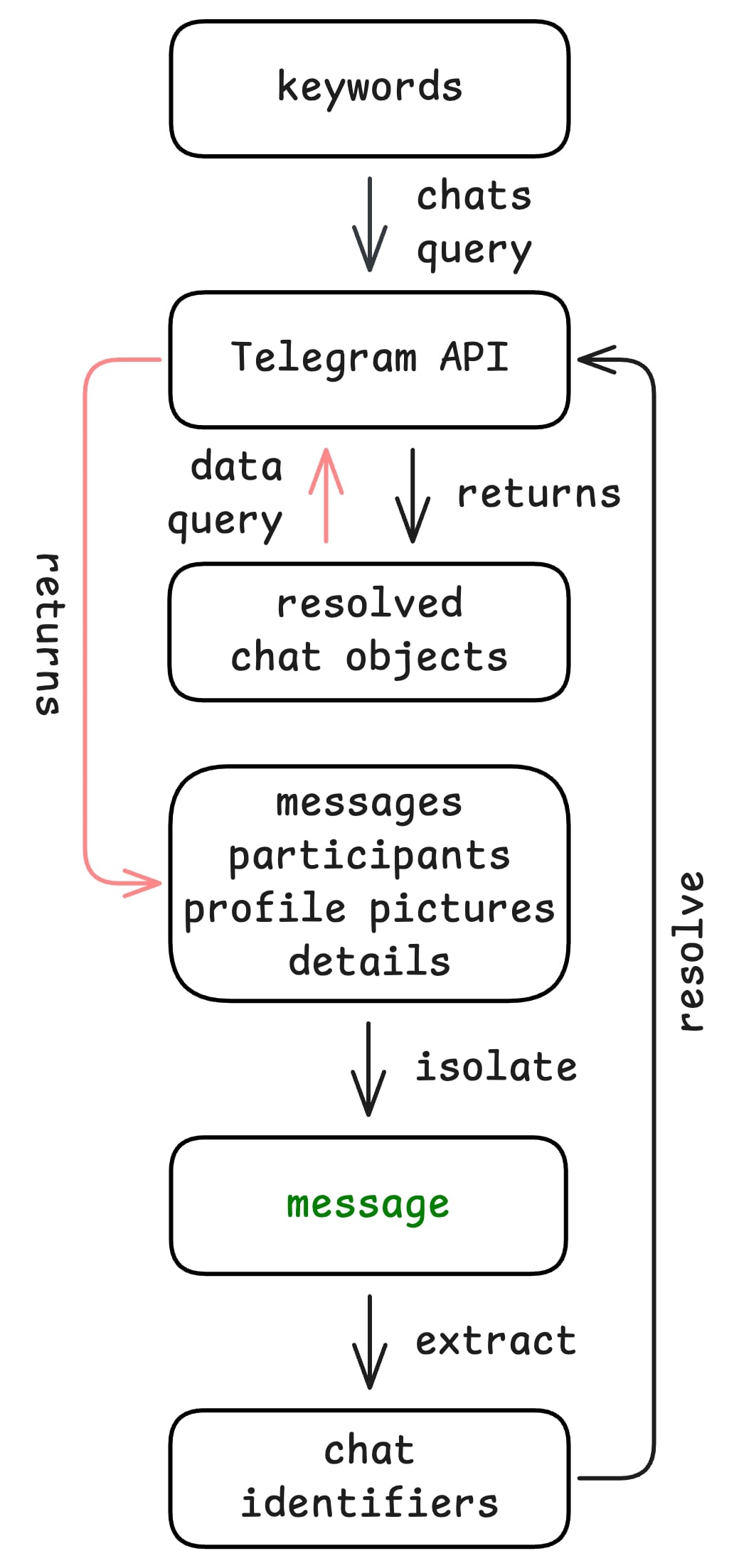}
		\caption{Data collection flow.}
		\vspace{-5\baselineskip}
		\label{fig:datacollectionflow}
	\end{wrapfigure}

 \section*{Data collection}
	\autoref{fig:datacollectionflow} outlines our data collection flow. The data was acquired by querying the Telegram API,\footnote{\url{https://core.telegram.org}} using the \inlinecode{telethon} library,\footnote{\url{https://docs.telethon.dev/en/stable}} for chats' details, profile pictures, messages, and participants' data. This process was seeded by searching a set of keywords (cf. Appendix) related to the election via the Telegram global search function,\footnote{\url{https://core.telegram.org/api/search}} to identify an initial set of publicly-accessible chats (henceforth, \textit{seed chats}) containing any of the searched keywords; next, every time that a link to another accessible chat was found in a \textit{seed chat}, that link was added to a list of \textit{to-be-scraped chats} --- the data collection, akin to a snowball sampling process, continued to recursively discover new chats and add them to the data collection.

 \smallskip \noindent{}The to-be-scraped chats' identifiers were discovered using the following criteria:

	\begin{itemize}
		\item Links, like "Tap here", leading to "\textbf{tpusa}.t.me".
		\item Text links, like "t.me/addlist/\textbf{\_iFZZv5diBY0OGEy}". 
		\item Mentions, like "@\textbf{trumptrain45}".
		\item Forwards, like "Forwarded from \textbf{joebidenchat}".
	\end{itemize}

 \smallskip \noindent{}Links containing chat identifiers were procured from "telegram.me/", "telegram.dog/", "t.me/", "<username>.t.me", and "tg://" links in \inlinecode{messageEntityTextUrl},\footnote{\url{https://core.telegram.org/constructor/messageEntityTextUrl}} and \newline \inlinecode{messageEntityUrl},\footnote{\url{https://core.telegram.org/constructor/messageEntityUrl}} objects. The processed links exposed:
	\begin{itemize}
		\item Usernames, like \textbf{jillsteincoin}.
		\item Group IDs, like channel \textbf{2036850729}.
		\item Chat invite codes,\footnote{\url{https://core.telegram.org/api/invites}} like \textbf{DOJPTuLa0DIxMDRk}.
		\item Chat folder codes,\footnote{\url{https://core.telegram.org/api/folders}} like \textbf{tSJoZOh1e503ZGJi}.
	\end{itemize}

 \clearpage
 
	Similarly, mentioned usernames were procured from \inlinecode{messageEntityMention} objects,\footnote{\url{https://core.telegram.org/constructor/messageEntityMention}} and forwarded usernames and channel IDs from \inlinecode{messageFwdHeader} objects.\footnote{\url{https://core.telegram.org/constructor/messageFwdHeader}} All four data types may be found in \inlinecode{message} objects.\footnote{\url{https://core.telegram.org/constructor/message}}
	
	Prior to scraping, the extracted usernames and invite codes were resolved through the Telegram API. Resolved usernames enabled public chat scraping, and resolved invite codes enabled public and private chat scraping. Resolving invite-based private chats, however, required chat membership, which was automatically requested using the invite codes. Analogously, folder codes underwent a resolve-like process---for simplicity, referred to as \textit{resolving}---which returned resolved private and public chats, enabling their scraping without membership requirements. Finally, chat IDs alone did not suffice to scrape, but served to track chats until usernames, invite codes, or folder codes were discovered. 
	
	\subsection*{Data collection considerations}
	Our collection was based on seed keywords, for setup searches, of at least 4 characters\footnote{\url{https://limits.tginfo.me/en}} (cf. Appendix). The chosen keywords represented relevant entities (candidates' names) or concepts (e.g., "election" or "debate"), and the occurrence of real-world events prompted us to track new keywords, like "Tim Walz", after he was announced as the Vice Presidential pick for the Democratic Party. Appendix A outlines the list of keywords and tracking start dates. Our goal was to record election-related chat textual data. Thus:
	\begin{itemize}
		\item Only message text---no files---sent on or after November 1, 2023, was collected.
		\item Only a participant's \inlinecode{user} object\footnote{\url{https://core.telegram.org/constructor/user}}---a distilled version of their full profile---was collected.
		\item Keywords not leading to relevant chats, like "Kamala2024", as of now, were not considered.
		\item The official Telegram channel, identified by the username \textbf{telegram}, was ignored.
	\end{itemize}
	 Still, not all scraped data is relevant---many messages advertise political meme coins, for instance---and, due to the scraper's recursive nature, it was unfeasible to scrape all discovered chats.
	
	The collection occurred from August 2, 2024, to October 31, 2024. Discovering and scraping new chats was prioritized until October 3, while a continuous update of the data set---adding new messages to already discovered chats---was prioritized from October 4. Throughout this period, some insurmountable situations occurred, including:
	\begin{itemize}
		\item Chats being deleted before, during, or after the scraping;
		\item Network situations interrupting a chat's scraping; 
		\item Server-side errors, rarely and specifically affecting chat profile pictures;
		\item Private chat administrators ignoring or rejecting join requests;
		\item Telegram banning scrapers---no reasons given---or chat administrators revoking chat access;
		\item New scrapers not gaining access and being unable to update private chats after a ban.
	\end{itemize}
	
	Similarly, given that Telegram usernames are reusable and exchangeable,\textsuperscript{\ref{note13}} and that public chats may have multiple usernames,\footnote{\url{https://telegram.org/faq\#usernames-and-t-me}\label{note13}} the collection frequently faced ambiguity. For instance, a message holding the \textbf{usavotes} identifier could have been referring to a chat that was banned or deleted, and not the discovered one. This was managed by assuming that the first discovered chat, if any, was the appropriate one, and by not conducting rediscoveries.
	
 Lastly, note that public user profiles and chats are accessible to all Telegram users. Likewise, private chats---not secret chats\footnote{\url{https://telegram.org/faq\#q-how-are-secret-chats-different}}---were discovered via invite and folder codes broadcasted in other chats, and no deceptive method was used to gain access to them: Join requests\footnote{\url{https://core.telegram.org/method/messages.importChatInvite}} do not support message exchanges, and no messages were sent to any user or bot.
	
	\FloatBarrier
	\begin{wrapfigure}{R}{0.40\textwidth}
		\includegraphics[width=0.40\textwidth]{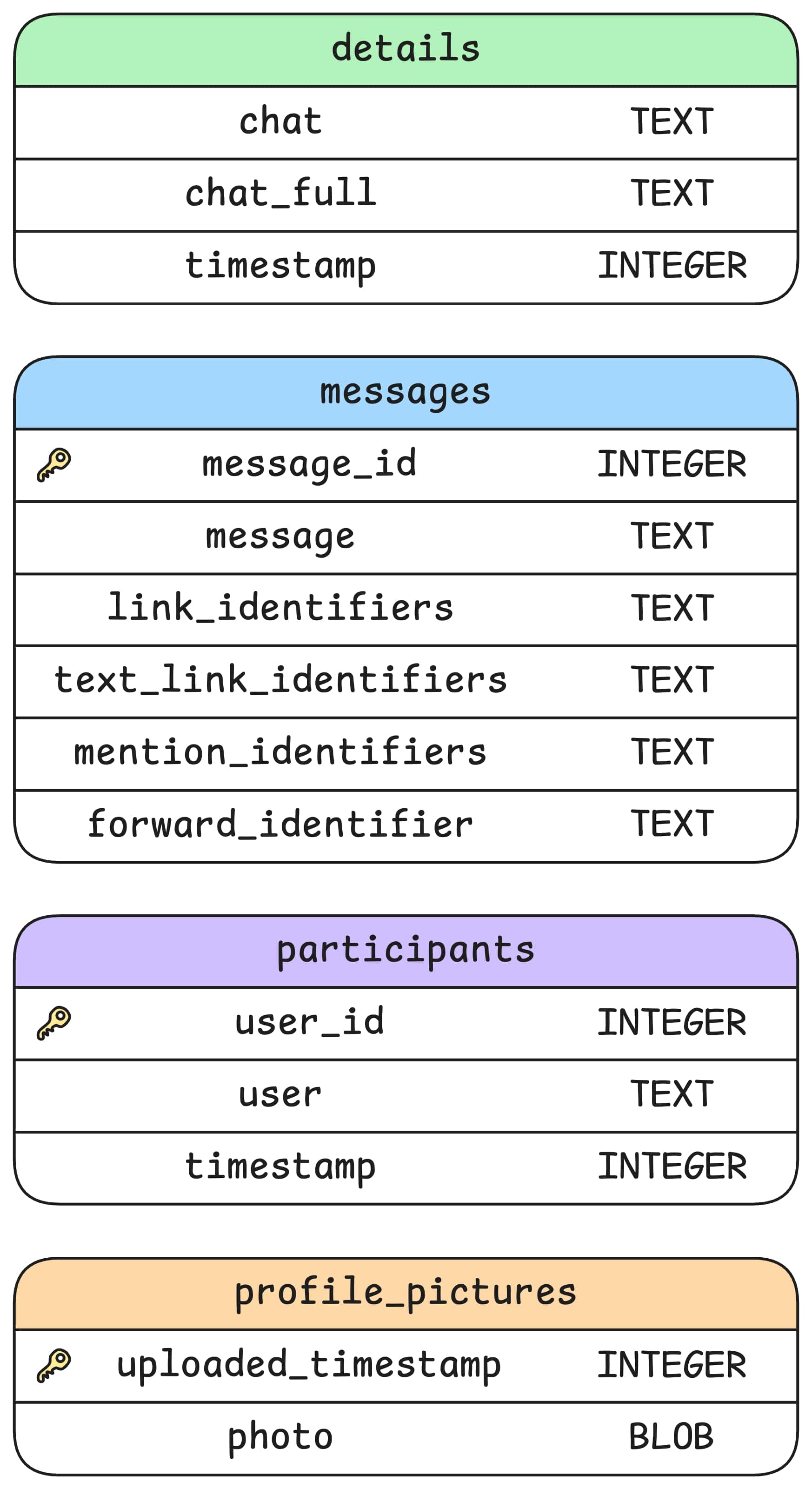}
		\label{fig:scrapedtables}
		\vspace{0.75\baselineskip}
		\includegraphics[width=0.40\textwidth]{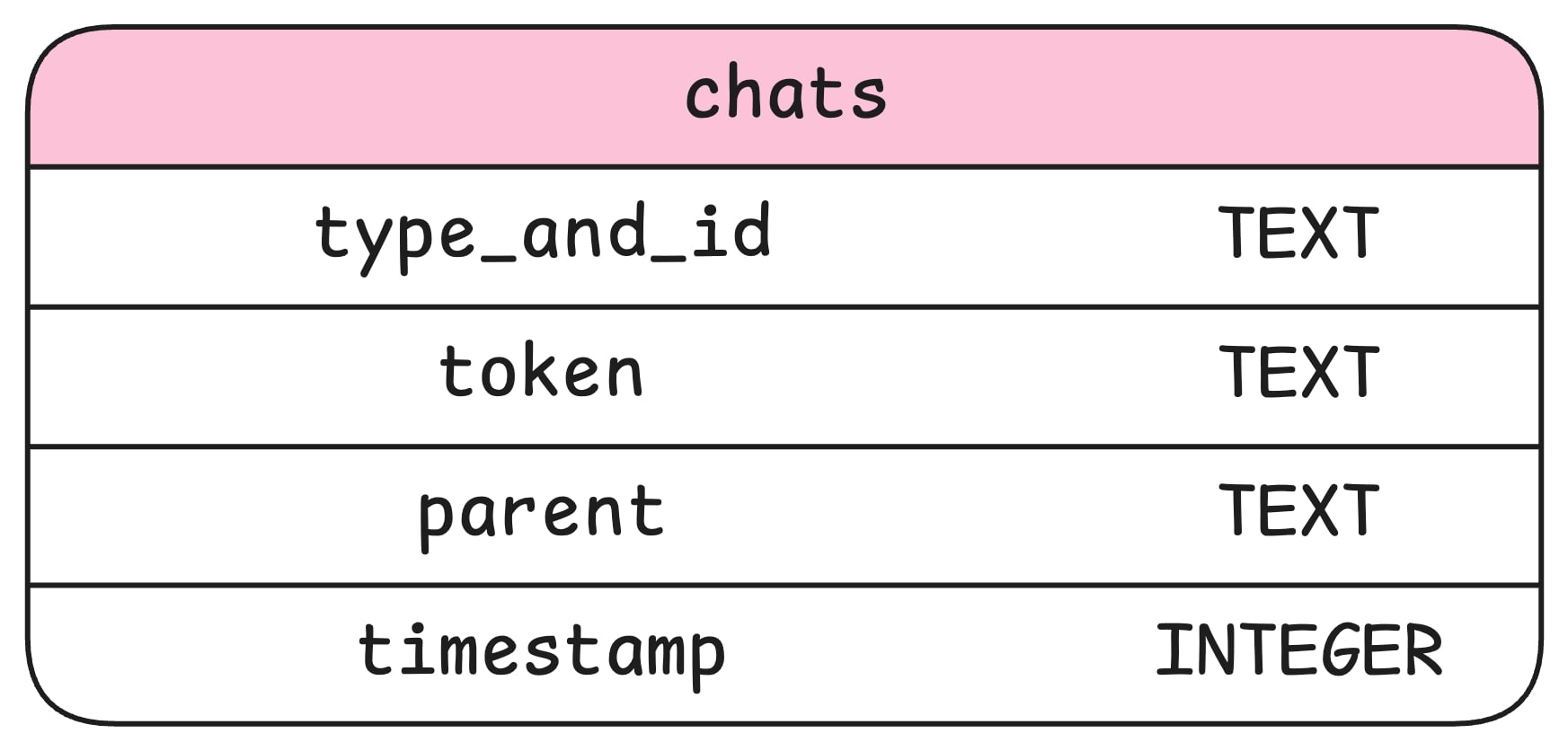}
		\caption{Our data schema.}
		\vspace{-1\baselineskip}
		\label{fig:schema}
	\end{wrapfigure}
	\FloatBarrier

	\section*{Data schema}
	A chat's data is stored in a SQLite database, "chat\_<chat ID>.db" or "channel\_<channel ID>.db". Each database hosts a "details" table, which stores \inlinecode{chat}\footnote{\url{https://core.telegram.org/constructor/chat}} or \inlinecode{channel}\footnote{\url{https://core.telegram.org/constructor/channel}} and \inlinecode{chatFull}\footnote{\url{https://core.telegram.org/constructor/chatFull}} or \inlinecode{channelFull}\footnote{\url{https://core.telegram.org/constructor/channelFull}} objects, and is timestamped at collection time. Additionally, each database may host:
	\begin{itemize}
		\item The "messages" table, storing \inlinecode{message} objects and relating them to their extracted chat identifiers. It is indexed using \inlinecode{message.id}, and timestamped via \inlinecode{message.date} or \inlinecode{message.edit_date}.
		\item The "participants" table, storing \inlinecode{user} objects, indexed using \inlinecode{user.id}, and timestamped at collection time.
		\item The "profile\_pictures" table, \inlinecode{photo} objects\footnote{\url{https://core.telegram.org/constructor/photo}} and indexed and timestamped using the time of upload, \inlinecode{photo.date}.
	\end{itemize}
	However, some of these tables may be absent if, at collection time, a chat has no: Messages sent on or after November 1, 2023; accessible participant data; or available profile pictures. Furthermore, \inlinecode{chat},  \inlinecode{channel}, \inlinecode{chatFull},  \inlinecode{channelFull}, \inlinecode{message}, and \inlinecode{user} objects were JSON-serialized, UTF-8 encoded, and compressed by using the \inlinecode{zlib.compress()}function.\footnote{\url{https://docs.python.org/3/library/zlib.html}} The tables' schema is illustrated in \autoref{fig:schema}.
	
	To complement the collected data, "chats", a SQLite table in "chats.db", is included. This table relates a chat's type and ID---if resolved or discovered---in the "chat\_<chat ID>" or "channel\_<channel ID>" formats, to its parents and tokens. A chat's parents are the chats, chat folders, and keywords which led to its discovering, and a chat's tokens are its usernames and invite codes. 
 
\clearpage
 The table "chats" has partial duplicates for different tokens or parents, is timestamped at resolve time, and its schema is illustrated in \autoref{fig:schema}.

	\subsection*{Data schema considerations}
	A chat in "chats" may have multiple or no tokens, as:
	\begin{itemize}
		\item Private chats do not have usernames.\footnote{\url{https://telegram.org/faq_channels\#q-how-are-public-and-private-channels-different}\label{note22}}
		\item Public chats must have a username,\textsuperscript{\ref{note22}} and may have multiple usernames.\textsuperscript{\ref{note13}}
		\item Public and private chats may have multiple invite codes.\footnote{\url{https://telegram.org/blog/autodelete-inv2\#expiring-invite-links}}
		\item Some invite codes, used to join private chats in past collections, were lost---which does not affect scraping or inter-chat relationship records.
	\end{itemize}
	Similarly, some tables, including "details", may be absent due to collection issues.
	
	Lastly, some extracted identifiers may be invalid due to semantic errors. For instance, a message containing the text link "t.me/c/\textbf{TartariaYoutube}"---which is invalid, as "t.me/c/" must be followed by a private channel ID,\footnote{\url{https://core.telegram.org/api/links}} ---results in the invalid "channel\_TartariaYoutube" identifier. Such identifiers did not impact the data collection or appear in "chats", but were allowed in "messages", as links with semantic errors may be valuable: At the time of writing, \textbf{tartariayoutube} leads to a channel. Note, however, that rectifying and scraping these identifiers was not attempted.
	
	
	\subsection*{Chat networks}
\begin{figure}[H]
		\centering
		\includegraphics[width=0.85\textwidth]{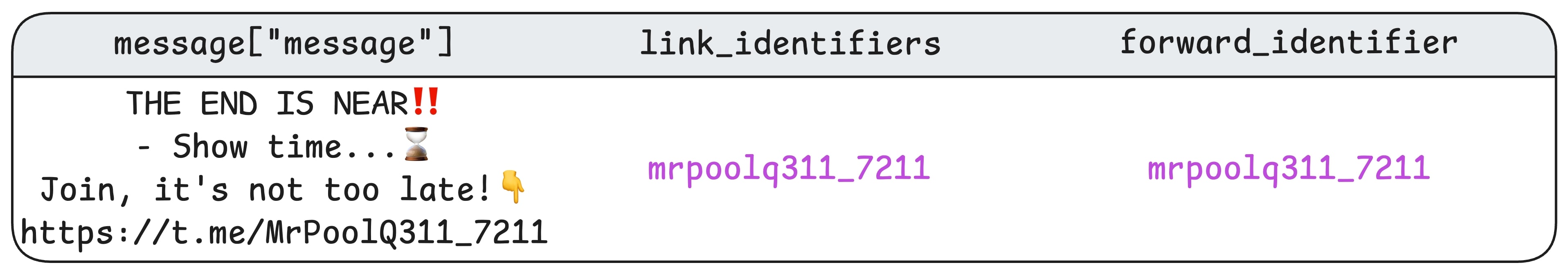}
		\caption{A message in channel 1516405535, connecting it to channel 2059913145.}
		\label{fig:chatmessage}
	\end{figure}

\begin{figure}[H]
		\centering
		\includegraphics[width=0.522682\textwidth]{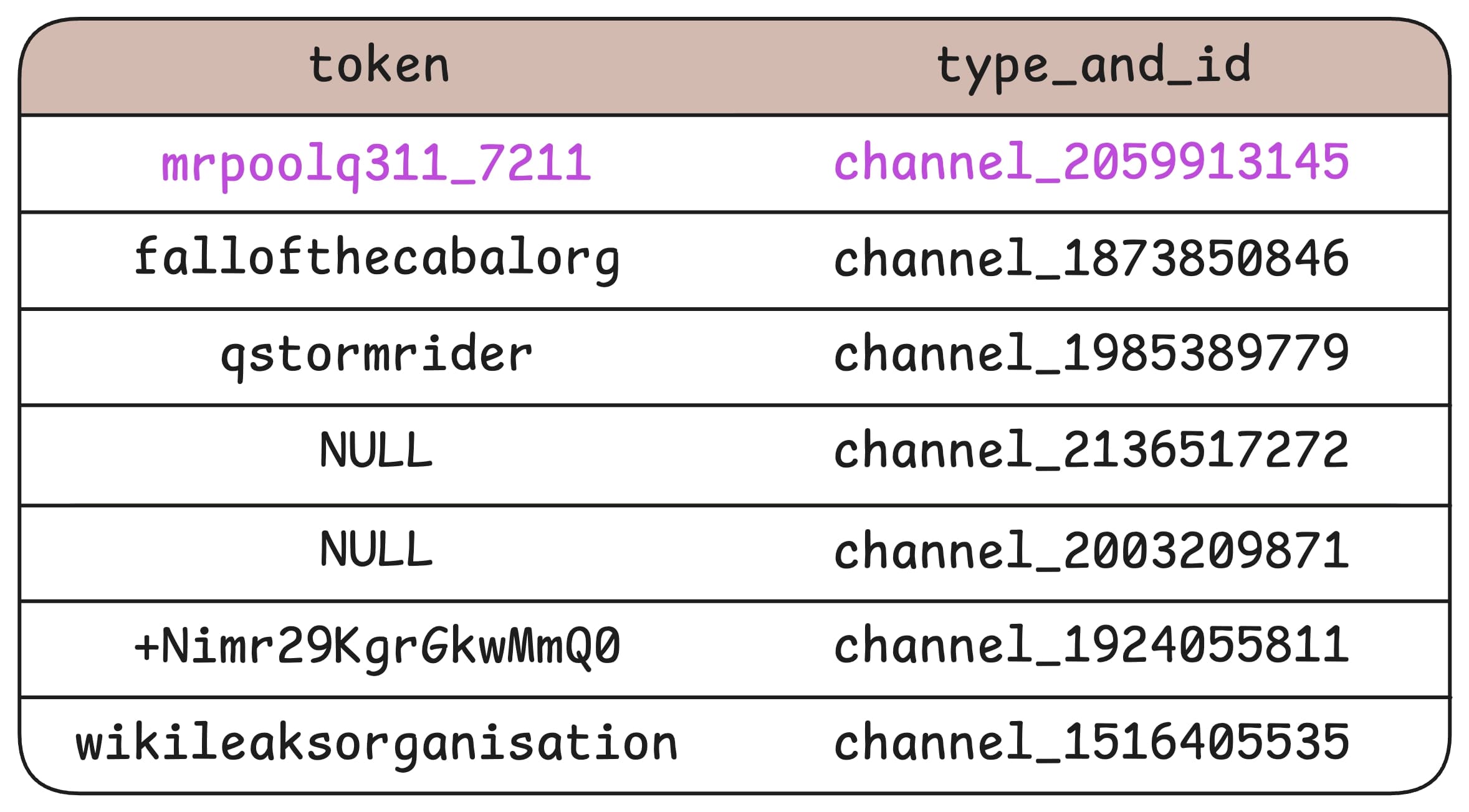}
		\caption{The chat identifiers extracted from channel 1516405535's messages, as of October 3, 2024.}
		\label{fig:chatrelationshipstable}
	\end{figure}

\begin{figure}[H]
		\centering
		\includegraphics[width=1\textwidth]{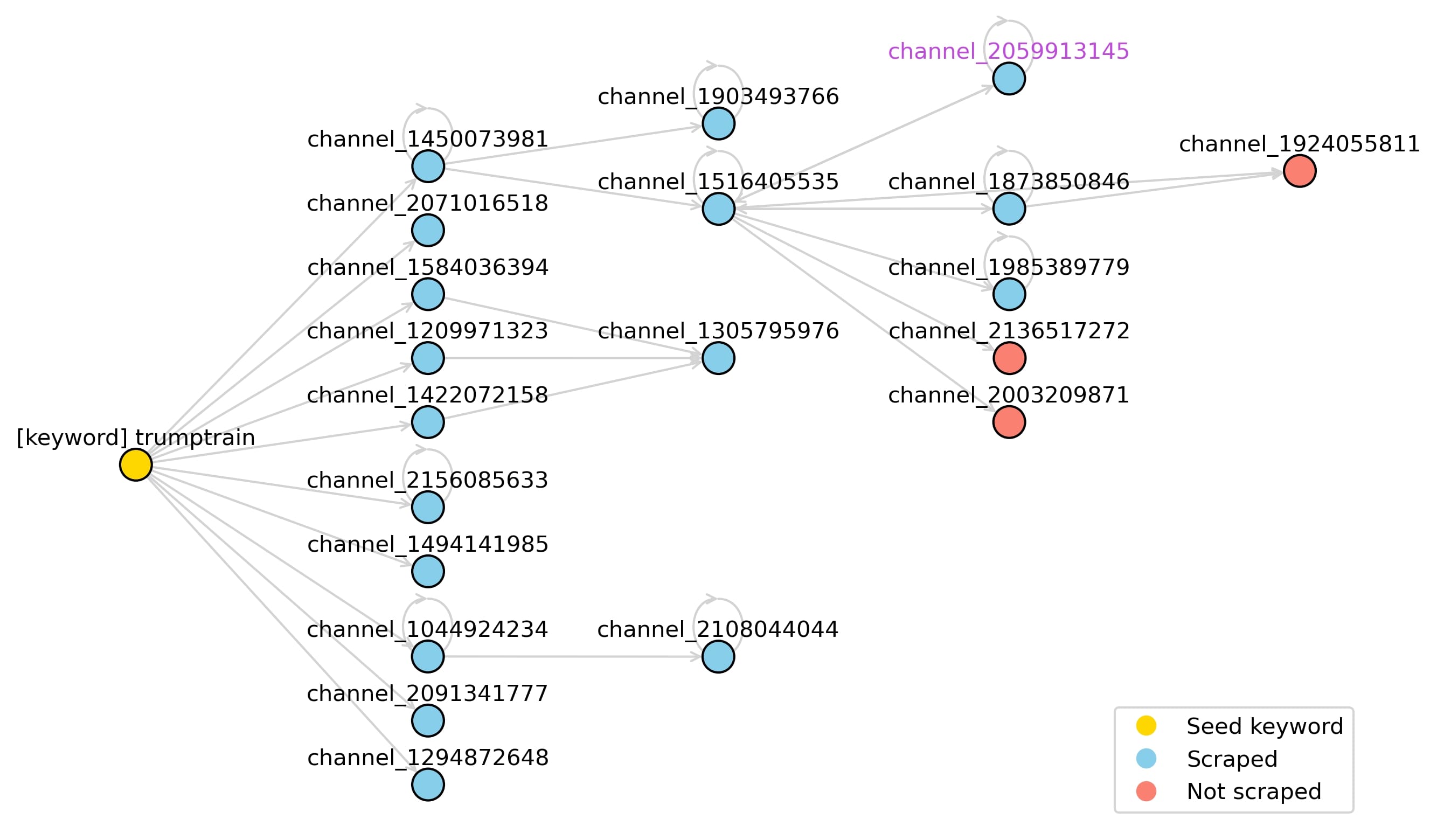}
		\caption{A subnetwork of resolved or discovered chats, as of October 3, 2024, containing channel 2059913145. Displaying three levels of recursive connections and unique directed edges per direction per vertex pair. Note that only chats in the dataset are considered scraped.}
		\label{fig:chatrelationshipsgraph}
	\end{figure}

	It is worth noting that the "messages" table also serves to understand the types of chat connections, as illustrated in the example in \autoref{fig:chatmessage}. The "chats" and "messages" tables serve to summarize chat connections, as illustrated in \autoref{fig:chatrelationshipstable}. Finally, "chats" serve to track inter-chat relationships, as shown in \autoref{fig:chatrelationshipsgraph}.

	\begin{wraptable}{r}{0.425\textwidth}
		\centering
		\renewcommand{\arraystretch}{0.785}
		\begin{tabular}{@{}c c@{}}
			\textbf{Language} & \textbf{Percentage} \\ \midrule
			English      & 57.72  \\
			Russian      & 14.24  \\
			German       & 1.08   \\
			Norwegian    & 0.53   \\
			Dutch        & 0.50   \\
			Afrikaans    & 0.24   \\
			Danish       & 0.23   \\
			Romanian     & 0.23   \\
			French       & 0.22   \\
			Italian      & 0.21   \\
			Somali 	& 0.21 \\
			Indonesian	& 0.21 \\
			Portuguese	& 0.19 \\
			Tagalog	& 0.18\\
			Welsh	& 0.14 \\
			Spanish	& 0.14 \\
			Catalan	& 0.14 \\
			Vietnamese	& 0.11 \\
			Estonian	& 0.1 \\
			Swahili	& 0.08 \\
		\end{tabular}
		\caption{The 20 most common languages in the 500 most influential chats.}
		\label{tab:languages}
	\end{wraptable}

	\section*{Dataset overview}
	The first release of our data collection features over 30 thousand chats and over 500 million messages---the largest public Telegram dataset by message count to the best of our knowledge. To provide a potentially meaningful dataset summarization and due to our dataset's sheer volume, the 500 most influential scraped chats---within the network of resolved or discovered chats---were selected for examination.
 
    To ascertain and quantify the influence of each chat in the network, we used the nodes' unique incoming edges; the top 500 chats, ranked by degree, 
 comprised 42,730,835 messages as of October 3, 2024.

 \subsection*{Exploratory Analysis}

By using the \inlinecode{langdetect} library,\footnote{\url{https://pypi.org/project/langdetect/}} we determined that English is the most prevalent language ($>$57\% of the messages), followed by Russian ($>$ 14\%). No language was identified in about 22\% of the messages, most of which exclusively contain emojis, URLs, or attachments. \autoref{tab:languages} lists the 20 most common languages.

	\clearpage
	
	\begin{wraptable}{r}{0.425\textwidth}
		\centering
		\renewcommand{\arraystretch}{0.785}
		\begin{tabular}{@{}c c@{}}
			\textbf{Word} & \textbf{Frequency} \\ \midrule
			trend         & 958,650            \\
			new           & 783,999            \\
			sol           & 760,364            \\
			market        & 687,132            \\
			cap           & 668,303            \\
			buy           & 657,113            \\
			x             & 499,159            \\
			profile       & 491,393            \\
			tg            & 487,268            \\
			nfa           & 487,215            \\
			dyor          & 487,175            \\
			web           & 313,765            \\
			people        & 263,780            \\
			say           & 258,295            \\
			chart         & 226,262            \\
			trump         & 226,225            \\
			group         & 219,617            \\
			event         & 187,068            \\
			time          & 185,675            \\
			know          & 172,564            \\
			force         & 172,289            \\
			year          & 170,687            \\
			russian       & 162,869            \\
			president     & 153,553            \\
			world         & 152,746            \\
			like          & 152,058            \\
			token         & 148,711            \\
			position      & 147,605            \\
			txn           & 147,596            \\
			join          & 146,626            \\
			day           & 146,466            \\
			ukraine       & 146,194            \\
			biden         & 146,091            \\
			israeli       & 142,351            \\
			military      & 142,285            \\
			state         & 142,222            \\
			gaza          & 140,575            \\
			go            & 139,921            \\
			israel        & 139,878            \\
			report        & 138,512            \\
			war           & 138,226            \\
			follow        & 135,009            \\
			attack        & 134,193            \\
			come          & 131,491            \\
			country       & 118,862            \\
			russia        & 116,894            \\
			break         & 114,946            \\
			al            & 111,057            \\
			ukrainian     & 110,921            \\
			good          & 109,862            \\
		\end{tabular}
		\vspace{1em}
		\caption{The 50 most common lemmas in the 500 most influential chats.}
		\label{tab:lemmas}
	\end{wraptable}

	
By using the \inlinecode{spacy} library,\footnote{\url{https://spacy.io/}} we extracted 66,591,042 lemmas from the English messages in the subset of analyzed chats. 
 \autoref{tab:lemmas} lists the 50 most common lemmas.
 Many of these relate to US politics, Russia, Ukraine, Gaza, and Israel. This indicates that, at least in the subset of analyzed chats, our dataset captured messages broadly relevant to the 2024 US Presidential Election and related geo-political discourse. Moreover, while seemingly irrelevant at first, online trading-related lemmas were often observed to belong in messages on political meme coins, like the "Trump Liberty Coin". For researchers who will use this dataset for analysis, we want to underscore the importance of appropriately filtering and cleaning the data to minimize the effects of the noise inherent to this type of keyword-based, large-scale data collections.
	
We further expanded this analysis to n-grams up to length 5 (i.e.,bigrams, trigrams, tetragrams, and pentagrams) that we extracted from the lemmatized English messages. The most common n-grams are related to spam and trading, however, some less-common sequences point to potentially problematic discussions, with possible several real-world implications, that are worthy of investigation. For instance, some messages include:
	\begin{itemize}
		\item "michigan create militia defend come"
		\item "new hampshire maga patriot come"
		\item "attack infrastructure vigilant fear patriot"
		\item "trade child sex"
		\item "sabotage maga"
	\end{itemize}
	
	Lastly, we extracted 1,305,353 URL domains from the English messages in the subset of analyzed chats. 
 \autoref{tab:domains} lists the 50 most common URL domains, where shortened domains were expanded, like t.co to x.com.
 Among these, domains from Telegram, social media platforms, and news and alternative news outlets were frequently observed. Note that captured Telegram links are processed as part of the recursive data collection and may lead to entries in the "chats" table.

	\clearpage
	
	\begin{wraptable}{r}{0.5\textwidth}
		\centering
		\renewcommand{\arraystretch}{0.785}
		\begin{tabular}{@{}c c@{}}
			\textbf{Domain} & \textbf{Frequency} \\ \midrule
			t.me                   & 381,490 \\ 
			x.com                  & 134,453 \\ 
			rumble.com             & 72,651 \\ 
			youtube.com            & 96,492 \\ 
			thegatewaypundit.com   & 30,157 \\ 
			iz.ru                  & 27,519 \\ 
			dailymail.co.uk        & 15,114 \\ 
			theepochtimes.com      & 10,352 \\ 
			truthsocial.com        & 9,411 \\ 
			substack.com           & 8,702 \\ 
			nypost.com             & 8,148 \\ 
			tiktok.com             & 8,053 \\ 
			resistthemainstream.com & 7,687 \\ 
			zerohedge.com          & 7,486 \\ 
			vigilantnews.com       & 7,378 \\ 
			bitchute.com           & 7,300 \\ 
			ept.ms                 & 7,296 \\ 
			msn.com                & 7,232 \\ 
			bit.ly                 & 7,201 \\ 
			facebook.com           & 7,099 \\ 
			naturalnews.com        & 7,067 \\ 
			rt.com                 & 7,015 \\ 
			thepostmillennial.com   & 7,005 \\ 
			insiderpaper.com       & 6,794 \\ 
			instagram.com          & 6,568 \\ 
			disclose.tv            & 5,846 \\ 
			odysee.com             & 5,653 \\ 
			isrefer.com            & 5,206 \\ 
			oann.com               & 4,796 \\ 
			reddit.com             & 5,161 \\ 
			yahoo.com              & 4,319 \\ 
			fxtwitter.com          & 4,234 \\ 
			foxnews.com            & 4,226 \\ 
			express.co.uk          & 4,157 \\ 
			fightback.law          & 4,075 \\ 
			amg-news.com           & 3,944 \\ 
			ruptly.tv              & 3,444 \\ 
			thenationalpulse.com    & 3,348 \\ 
			theguardian.com        & 3,185 \\ 
			livefree.be            & 3,148 \\ 
			reuters.com            & 2,962 \\ 
			fenbenlab.com          & 2,774 \\ 
			google.com             & 2,697 \\ 
			breitbart.com          & 2,647 \\ 
			wikipedia.org          & 2,430 \\ 
			tinyurl.com            & 2,410 \\ 
			southfront.press       & 2,405 \\ 
			lawenforcementtoday.com & 2,211 \\ 
			modernity.news         & 2,201 \\ 
			cnn.com                & 2,138 \\ 
		\end{tabular}
		\vspace{1em}
		\caption{The 50 most common URL domains in the 500 most influential chats.}
		\label{tab:domains}
	\end{wraptable}
 
	\section*{Conclusions}
	In this paper, we presented a large-scale Telegram dataset focused on the 2024 US  Presidential Election, which we are making public–––the largest Telegram data release to our knowledge. 
	Our first data release features over 30 thousand chats and over half a billion messages, and includes chats' details, profile pictures, messages, and participant information. The collection relied on the Telegram API and implemented a recursive scraping process using the chat identifiers discovered in messages' forwards, mentions, links and text links. The first channels were discovered via global searches, using seed keywords, and the collected data included a chat's details, profile pictures, messages, and participant information. Chat network relationships were captured and used as proxy to infer a ranking of chats' importance: for sake of preliminary analysis, the top 500 chats were selected for examination. The analyses suggested that the selected messages:
	\begin{itemize}
		\item Are mostly in English or Russian.
		\item Often contain lemmas related to US politics, Russia, Ukraine, Gaza, Israel, and online trading.
		\item Frequently contain links to social platforms, news outlets, and alternative news media. 
		\item Occasionally contain text hinting to potentially concerning, problematic, or harmful ideas.
	\end{itemize}

 \section*{Dataset Access}

 \subsection*{Release v1 (Oct 31, 2024)}
 The first dataset's publicly release (v1) contains data collected from August 2, 2024, to October 31, 2024. This dataset is still being continuously collected and will be routinely updated.

 The dataset and all ancillary information are available at the following Github repository: 
 \url{https://github.com/leonardo-blas/usc-tg-24-us-election}

\clearpage
	\section*{About the Team}
	The 2024 Election Integrity Initiative is led by Emilio Ferrara and Luca Luceri and carried out by a collective of USC students and volunteers whose contributions are instrumental to enable these studies. The authors are indebted to Srilatha Dama and Zhengan Pao for their help in bootstrapping this data collection. The authors are also grateful to the following HUMANS Lab's members for their tireless efforts on this project: Ashwin Balasubramanian, Leonardo Blas, Charles 'Duke' Bickham, Keith Burghardt, Sneha Chawan, Vishal Reddy Chintham, Eun Cheol Choi, Priyanka Dey, Isabel Epistelomogi, Saborni Kundu, Grace Li, Richard Peng, Gabriela Pinto, Jinhu Qi, Ameen Qureshi, Tanishq Salkar, Kashish Atit Shah, Reuben Varghese, Siyi Zhou.
\textbf{Previous memos}: \cite{memo1,memo2,memo3,memo4,memo6,memo7}

\bibliographystyleMemo{unsrt}
\bibliographyMemo{memos}

\bibliographystyle{ACM-Reference-Format}
\bibliography{memos}

\newpage
\section*{A \quad Keywords}
\begin{table}[H]
    \centering
    \renewcommand{\arraystretch}{0.9}
    \begin{tabular}{c c}
        \textbf{Keyword} & \textbf{Searched on} \\ \midrule
        2024 Elections & 8/2/24 \\
        2024 Presidential Elections & 8/2/24 \\
        2024 US Elections & 8/2/24 \\
        assassination & 8/2/24 \\
        Biden & 8/2/24 \\
        Biden2024 & 8/2/24 \\
        bidenharris2024 & 8/2/24 \\
        conservative & 8/2/24 \\
        Cornel West & 8/2/24 \\
        Dean Phillips & 8/2/24 \\
        Democratic party & 8/2/24 \\
        democratsoftiktok & 8/2/24 \\
        Donald Trump & 8/2/24 \\
        Independent Party & 8/2/24 \\
        JD Vance & 8/2/24 \\
        Jill Stein & 8/2/24 \\
        Joe Biden & 8/2/24 \\
        Joseph Biden & 8/2/24 \\
        Kamala Harris & 8/2/24 \\
        letsgobrandon & 8/2/24 \\
        MAGA & 8/2/24 \\
        makeamericagreatagain & 8/2/24 \\
        Marianne Williamson & 8/2/24 \\
        Nikki Haley & 8/2/24 \\
        phillips2024 & 8/2/24 \\
        Republican party & 8/2/24 \\
        republicansoftiktok & 8/2/24 \\
        RFK Jr & 8/2/24 \\
        Robert F. Kennedy Jr. & 8/2/24 \\
        Ron DeSantis & 8/2/24 \\
        thedemocrats & 8/2/24 \\
        Third Party & 8/2/24 \\
        Trump2024 & 8/2/24 \\
        trumpsupporters & 8/2/24 \\
        trumptrain & 8/2/24 \\
        ultramaga & 8/2/24 \\
        US Elections & 8/2/24 \\
        Vivek Ramaswamy & 8/2/24 \\
        voteblue2024 & 8/2/24 \\
        williamson2024 & 8/2/24 \\
        Tim Walz & 8/6/24 \\
    \end{tabular}
    \vspace{1em}
    \caption{Keywords and search dates.}
    \label{tab:electiontermsdates}
\end{table}

\end{document}